\documentstyle[11pt,epsfig]{article}
\begin{document}
\title{\bf FRW Cosmology From Five Dimensional Vacuum Brans--Dicke Theory}
\author{Amir F. Bahrehbakhsh\thanks{email: af-bahrehbakhsh@sbu.ac.ir},\, Mehrdad Farhoudi\thanks{email:
 m-farhoudi@sbu.ac.ir} \ and Hossein Shojaie\thanks{email: h-shojaie@sbu.ac.ir}\\
 {\small Department of Physics, Shahid Beheshti University, G.C., Evin, Tehran 19839, Iran}}
\date{\small September 17, 2010}
\maketitle
\begin{abstract}
We follow the approach of induced--matter theory for a
five--dimensional ($5D$) vacuum Brans--Dicke theory and introduce
induced--matter and induced potential in four dimensional ($4D$)
hypersurfaces, and then employ a generalized FRW type solution. We
confine ourselves to the scalar field and scale factors be
functions of the cosmic time. This makes the induced potential, by
its definition, vanishes, but the model is capable to expose
variety of states for the universe. In general situations, in
which the scale factor of the fifth dimension and scalar field
are~not constants, the $5D$ equations, for any kind of geometry,
admit a power--law relation between the scalar field and scale
factor of the fifth dimension. Hence, the procedure exhibits that
$5D$ vacuum FRW--like equations are equivalent, in general, to the
corresponding $4D$ vacuum ones with the same spatial scale factor
but a new scalar field and a new coupling constant,
$\tilde{\omega}$. We show that the $5D$ vacuum FRW--like
equations, or its equivalent $4D$ vacuum ones, admit accelerated
solutions. For a constant scalar field, the equations reduce to
the usual FRW equations with a typical radiation dominated
universe. For this situation, we obtain dynamics of scale factors
of the ordinary and extra dimensions for any kind of geometry
without any \emph{priori} assumption among them. For non--constant
scalar fields and spatially flat geometries, solutions are found
to be in the form of power--law and exponential ones. We also
employ the weak energy condition for the induced--matter, that
gives two constraints with negative or positive pressures. All
types of solutions fulfill the weak energy condition in different
ranges. The power--law solutions with either negative or positive
pressures admit both decelerating and accelerating ones. Some
solutions accept a shrinking extra dimension. By considering
non--ghost scalar fields and appealing the recent observational
measurements, the solutions are more restricted. We illustrate
that the accelerating power--law solutions, which satisfy the weak
energy condition and have non--ghost scalar fields, are compatible
with the recent observations in ranges $-4/3<\omega\leq-1.3151$
for the coupling constant and $1.5208\leq n<1.9583$ for dependence
of the fifth dimension scale factor with the usual scale factor.
These ranges also fulfill the condition $\tilde{\omega}>-3/2$
which prevents ghost scalar fields in the equivalent $4D$ vacuum
Brans--Dicke equations. The results are presented in a few tables
and figures.
\end{abstract}
\bigskip
\section{Introduction}
\indent

Attempts to geometrical unification of gravity with other
interactions, using higher dimensions other than our conventional
$4D$ space--time, began shortly after invention of the special
relativity (\textbf{SR}). Nordstr{\o}m was the first who built a
unified theory on the base of extra dimensions~\cite{1}. Tight
connection between SR and electrodynamics, namely the Lorentz
transformation, led Kaluza~\cite{2} and Klein~\cite{3} to
establish $5D$ versions of general relativity (\textbf{GR}) in
which electrodynamics rises from the extra fifth dimension. Since
then, considerable amount of works have been focused on this idea
either using different mechanism for compactification of extra
dimension or generalizing it to non--compact scenarios (see e.g.
Ref.~\cite{OverduinWesson1997}) such as Brane--World
theories~\cite{4}, space--time--matter or induced--matter
(\textbf{IM}) theories~\cite{5} and references therein. The latter
theories are based on the Campbell--Magaard theorem which asserts
that any analytical $N$--dimensional Riemannian manifold can
locally be embedded in an $(N+1)$--dimensional Ricci--flat
Riemannian manifold~\cite{6}. This theorem is of great importance
for establishing $4D$ field equations with matter sources locally
to be embedded in $5D$ field equations without \emph{priori}
introducing matter sources. Indeed, the matter sources of $4D$
space--times can be viewed as a manifestation of extra dimensions.
This is actually the core of IM theory which employs GR as the
underlying theory.

On the other hand, Jordan~\cite{9} attempted to embed a curved
$4D$ space--time in a flat $5D$ space--time and introduced a new
kind of gravitational theory, known as the scalar--tensor theory.
Following his idea, Brans and Dicke~\cite{10} invented an
attractive version of the scalar--tensor theory, an alternative to
GR, in which the weak equivalence principle is saved and a
non--minimally scalar field couples to curvature. The advantage of
this theory is that it is more Machian than GR, though mismatching
with the solar system observations is claimed as its
weakness~\cite{Bertotti2003fujiBook}. However, the solar system
constraint is a generic difficulty in the context of the
scalar--tensor theories~\cite{banerjee2001Sen2001}, and it
does~not necessarily denote that the evolution of the universe, at
all scales, should be close to GR, in which there are some debates on its tests on
cosmic scales~\cite{bean2009}.

Although it is sometimes desirable to have a higher dimensional
energy--momentum tensor or a scalar field, for example in
compactification of extra curved dimensions~\cite{11}, but the
most preference of higher dimensional theories is to obtain
macroscopic $4D$ matter from pure geometry. In this approach, some
features of a $5D$ vacuum Brans--Dicke (\textbf{BD}) theory based
on the idea of IM theory have recently been
demonstrated~\cite{12}, in where the role of GR as fundamental
underlying theory has been replaced by the BD theory of
gravitation. Actually, it has been shown that $5D$ vacuum BD
equations, when reduced to four dimensions, lead to a modified
version of the $4D$ Brans--Dicke theory which includes an induced
potential. Whereas in the literature, in order to obtain
accelerating universes, inclusion of such potentials has been
considered in \emph{priori} by hand. A few applications and a
$D$--dimensional version of this approach have been
performed~\cite{Ponce1and2,12.5}. Though, in
Refs.~\cite{Ponce1and2}, it has also been claimed that their
procedure provides explicit definitions for the effective matter
and induced potential. Besides, some misleading statements and
equations have been asserted in Ref.~\cite{12}, and hence we have
re--derived the procedure in Section $2$. Actually, the reduction
procedure of a $5D$ analogue of the BD theory, with matter
content, on every hypersurface orthogonal to an extra cyclic
dimension (recovering a modified BD theory described by a
4--metric coupled to two scalar fields) has previously been
performed in the literature~\cite{qiang20052009}. However, the key
point of IM theories are based on not introducing matter sources
in $5D$ space--times.

In addition, recent measurements of anisotropies in the microwave
background suggest that our ordinary $4D$ universe should be
spatially flat~\cite{13}, and the observations of Type Ia--supernovas
indicate that the universe is in an accelerating expansion
phase~\cite{14}. Hence, the universe should mainly be filled with
a dark energy or a quintessence which makes it to expand with
acceleration~\cite{15}. Then after an intensive amount of work has been
performed in the literature to explain the acceleration of the
universe.

In this work, we explore the Friedmann--Robertson--Walker
(\textbf{FRW}) type cosmology of a $5D$ vacuum BD theory and
obtain solutions and related conditions. This model has extra
terms, such as a scalar field and scale factor of fifth dimension,
which make it capable to present accelerated universes beside
decelerated ones. In the next section, we give a brief review of
the induced modified BD theory from a $5D$ vacuum space--time to
rederive the induced energy--momentum tensor, as has been
introduced in Ref.~\cite{12}, for our purpose to employ the energy
density and pressure. In Section~$3$, we consider a generalized
FRW metric in the $5D$ space--time and specify FRW cosmological
equations and employ the weak energy condition (\textbf{WEC}) to
obtain the energy density and pressure conditions. Then, we probe
two special cases of a constant scale factor of the fifth
dimension and a constant scalar field. In Section~$4$, we proceed
to exhibit that $5D$ vacuum BD equations, employing the
generalized FRW metric, are equivalent, in general, to the
corresponding vacuum $4D$ ones. This equivalency can be viewed as
the main point within this work which distinguishes it from
Refs.~\cite{12,Ponce1and2}. In Section~$5$, we find exact
solutions for flat geometries and proceed to get solutions
fulfilling the WEC while being compatible with the recent
observational measurements. We also provide a few tables and
figures for a better view of acceptable range of parameters.
Finally, conclusions are presented in the last section.

\section{Modified Brans--Dicke Theory From Five--Dimensional Vacuum}
\indent

Following the idea of IM theories~\cite{5}, one can replace GR by
the BD theory of gravitation as the underlying
theory~\cite{12,Ponce1and2,qiang20052009}. For this purpose, the
action of $5D$ Brans--Dicke theory can analogously be written in
the Jordan frame as
\begin{equation}\label{1}
\emph{S} \\\
[g_{_{AB}},\phi]=\int\sqrt{|{}^{_{(5)}}g|} \left (\phi \
^{^{(5)}}\!R-\frac{\omega}{\phi}g^{_{AB}}\phi_{,_{A}}\phi_{,_{B}}+
16\pi L_{m} \right )d^{5}x\, ,
\end{equation}
where $c=1$, the capital Latin indices run from zero to four,
$\phi$ is a positive scalar field that describes gravitational coupling in
five dimensions, $^{^{(5)}}R$ is $5D$ Ricci scalar, $^{_{(5)}}g$
is the determinant of $5D$ metric $g_{_{AB}}$, $\emph{L}_{m}$
represents the matter Lagrangian and $\omega$ is a dimensionless
coupling constant. The field equations obtained from action (1)
are
\begin{equation}\label{2}
^{^{(5)}}G_{_{AB}}=\frac{8\pi}{\phi} \
^{^{(5)}}T_{_{AB}}+\frac{\omega}{\phi^2}
\left(\phi_{,_{A}}\phi_{,_{B}}-\frac{1}{2}g_{_{AB}}\phi^{,_{C}}\phi_{,_{C}}
\right)+\frac{1}{\phi}\left(\phi_{;_{AB}}-g_{_{AB}}\
^{^{(5)}}\Box\phi \right)
\end{equation}
and
\begin{eqnarray}\label{3}
^{^{(5)}}\Box\phi=\frac{8\pi}{4+3\omega}\ ^{^{(5)}}T\, ,
\end{eqnarray}
where $^{^{(5)}}\Box\! \equiv{}_{;_{A}}{}^{^{A}}$,
$^{^{(5)}}G_{_{AB}}$ is $5D$ Einstein tensor ,
$^{^{(5)}}T_{_{AB}}$ is $5D$ energy--momentum tensor, $^{^{(5)}}T
\equiv\ ^{^{(5)}}T^{^{C}}{} _{^{C}}$. Also, in order to have a
non--ghost scalar field in the conformally related Einstein frame,
i.e. a field with a positive kinetic energy term in that frame,
the BD coupling constant must be
$\omega>-4/3$~\cite{Freund1982,16}.

As explained in the introduction, we propose to consider a $5D$
vacuum state, i.e. $^{^{(5)}}T_{_{AB}}= 0 =\ ^{^{(5)}}T$, where
equations (2) and (3) read
\begin{equation}\label{4}
^{^{(5)}}G_{_{AB}}=\frac{\omega}{\phi^2}\left(\phi_{,_{A}}\phi_{,_{B}}-\frac{1}{2}
g_{_{AB}}\phi^{,_{C}}\phi_{,_{C}}\right)
+\frac{1}{\phi}\left(\phi_{;_{AB}}-g_{_{AB}}\ ^{^{(5)}}\Box\phi\right)
\end{equation}
and\footnote{We have purposely kept the null term in equation (\ref{4}) for later on convenient.}
\begin{equation}\label{5}
^{^{(5)}}\Box\phi=0.
\end{equation}
For cosmological purposes one usually restricts attention to $5D$
metrics of the form, in local coordinates $x^{A}=(x^{\mu},y)$,
\begin{equation}\label{6}
dS^2=g_{_{AB}}(x^{C})dx^{A}dx^{B}=\
^{^{(5)}}\!g_{\mu\nu}(x^{C})dx^{\mu}dx^{\nu}+g_{_{44}}(x^{C})dy^2
\equiv \ ^{^{(5)}}\!g_{\mu\nu}(x^{C})dx^{\mu}dx^{\nu}+\epsilon
b^{2}(x^{C})dy^2\, ,
\end{equation}
where $y$ represents the fifth coordinate, the Greek indices run
from zero to three and $\epsilon^2=1$. It should be noted that
this ansatz is restrictive, but one limits oneself to it for
reasons of simplicity. Assuming the $5D$ space--time is foliated
by a family of hypersurfaces, $\Sigma$, defined by fixed values of
the fifth coordinate, then the metric intrinsic to every generic
hypersurface, e.g. $\Sigma_{o}(y=y_{o})$, can be obtained when
restricting the line element (\ref{6}) to displacements confined
to it. Thus, the induced metric on the hypersurface $\Sigma_{o}$
can have the form
\begin{equation}\label{6.1}
ds^2=\
^{^{(5)}}\!g_{\mu\nu}(x^{\alpha},y_{o})dx^{\mu}dx^{\nu}\equiv
g_{\mu\nu}dx^{\mu}dx^{\nu}\, ,
\end{equation}
in such a way that the usual $4D$ space--time metric, $g_{\mu\nu}$, can be recovered.

Hence, equation (\ref{4}) on the hypersurface $\Sigma_{o}$ can be
written as
\begin{equation}\label{6.4a}
G_{\alpha\beta}=\frac{8\pi}{\phi}T^{^{\rm(BD)}}_{\alpha\beta}+\frac{\omega}{\phi^2}
 \Big (\phi_{,\alpha}\phi_{,\beta}-\frac{1}{2}g_{\alpha\beta}\phi^{,\sigma}\phi_{,\sigma} \Big )
+\frac{1}{\phi}\bigg [\phi_{;\alpha\beta}-g_{\alpha\beta}\Big (\Box\phi-\frac{1}{2}V(\phi)\Big )\bigg ],
\end{equation}
where $T^{^{\rm(BD)}}_{\alpha\beta}$ is an induced
energy--momentum tensor of the effective $4D$ modified BD theory,
which is defined as
\begin{equation}\label{6.4}
T^{^{\rm(BD)}}_{\alpha\beta}\equiv T^{^{\rm(IM)}}_{\alpha\beta}+T^{^{(\phi)}}_{\alpha\beta}\, ,
\end{equation}
with\footnote{We have corrected miscalculations mentioned in the
Introduction.}
\begin{eqnarray}\label{6.5}
T^{^{\rm(IM)}}_{\alpha\beta}\equiv\frac{\phi}{8\pi}
\Bigg \{  \frac{b_{;\alpha\beta}}{b}-\frac{\Box b}{b}g_{\alpha\beta}-\frac{\epsilon}{2b^2}
\Bigg [\frac{b'}{b}g'_{\alpha\beta} -g''_{\alpha\beta}+g^{\mu\nu}g'_{\alpha\mu}
g'_{\beta\nu}-\frac{1}{2}g^{\mu\nu}g'_{\mu\nu}g'_{\alpha\beta}
\nonumber\\
-g_{\alpha\beta}\bigg (\frac{b'}{b}g^{\mu\nu}g'_{\mu\nu}-g^{\mu\nu}g''_{\mu\nu}-\frac{1}
{4}g^{\mu\nu}g^{\rho\sigma}g'_{\mu\nu}g'_{\rho\sigma}
-\frac{3}{4}g'^{\mu\nu}g'_{\mu\nu} \bigg )\Bigg ]
\Bigg \}
\end{eqnarray}
and
\begin{equation}\label{6.6}
T^{^{(\phi)}}_{\alpha\beta}\equiv-\frac{\epsilon}{8\pi b^2}\Bigg
\{g_{\alpha\beta}\Bigg [\phi''+\Big (\frac{1}{2}g^{\mu\nu}g'_{\mu\nu}-\frac{b'}{b}\Big )
\phi'+\epsilon bb_{,\mu}\phi^{,\mu} \Bigg ]-\frac{1}{2}g'_{\alpha\beta}\phi' \Bigg \}.
\end{equation}
Also, the induced potential has been defined in the formal identification as~\cite{12}
\begin{equation}\label{6.3}
V[\phi]\equiv -\epsilon \frac{\omega}{b^{2}}\frac{\phi^{'2}}{\phi} \Big |_{_{\Sigma_{0}}} ,
\end{equation}
where the prime denotes derivative with respect to the fifth coordinate.
Such an identification has been claimed~\cite{agular05anabitarte06} to be valid depending
on metric background and considering separable scalar fields.
However, this definition is different from what has been used in Ref.~\cite{Ponce1and2}.

Reduction of equation (\ref{5}) on the hypersurface $\Sigma_{o}$
gives
\begin{equation}\label{freduction}
\Box\phi=-\frac{\epsilon}{b^{2}}\left[\phi''+\phi'\left(\frac{g^{\alpha\beta}g'_{\alpha\beta}}{2}-\frac{b'}{b}\right)
\right]-\frac{b_{,\mu}}{b}\phi^{,\mu}\, ,
\end{equation}
which after manipulation resembles the other field equation of a
modified BD theory in four dimensions with induced potential. The
definition $T^{^{\rm(BD)}}_{\alpha\beta}$ and equation
(\ref{freduction}) are all we need for our purpose in this work
and an interested reader can consult Refs.~\cite{12,Ponce1and2}
for further details.

In the next section we assume a generalized FRW metric in a vacuum $5D$
universe to find its cosmological implications.

\section{Generalized FRW Cosmology}
\indent

For a $5D$ universe with an extra space--like dimension in addition to the three usual
spatially homogenous and isotropic ones, metric (\ref{6}) can be written as
\begin{equation}\label{7}
dS^2=-dt^2+a^2(t,y)\left [\frac{dr^2}{1-kr^2}+r^2(d\theta^2+\sin^2\theta
d\varphi^2)\right ]+b^2(t,y)dy^2\,,
\end{equation}
that can be considered as a generalized FRW solution. The scalar
field $\phi$ and the scale factors $a$ and $b$, in general, are
functions of $t$ and $y$. However, for simplicity and physical
plausibility, we assume the extra dimension is cyclic, i.e. the
hypersurface--orthogonal space--like is a Killing vector field in
the underlying $5D$ space--time~\cite{qiang20052009}. Hence, all
fields are functions of the cosmic time only, and definition
(\ref{6.3}) makes the induced potential vanishes. In this case, we
will show that such a universe can have accelerating and
decelerating solutions. Note that, the functionality of the scale
factor $b$ on $y$, either can be eliminated by transforming to a
new extra coordinate if $b$ is a separable function, and or makes
no changes in the following equations if $b$ is the only field
that depends on $y$. Besides, in the compactified extra dimension
scenarios, all fields are Fourier--expanded around $y_{o}$, and
henceforth one can have terms independent of $y$ to be observable,
i.e. physics would thus be effectively independent of compactified
fifth dimension~\cite{OverduinWesson1997}.

Considering metric (\ref{7}),
equations (\ref{4}) and (\ref{5}) result in cosmological equations
\begin{equation}\label{8}
H^2-\frac{\omega}{6}F^2+HF+\frac{k}{a^2}=-\left(HB+\frac{1}{3}BF\right),
\end{equation}
\begin{equation}\label{9}
2\dot{H}+\dot{F}+3H^2+\left(\frac{\omega}{2}+1\right)F^2+2HF+\frac{k}{a^2}=-\left(\dot{B}+B^2+2HB+BF\right),
\end{equation}
\begin{equation}\label{10}
2\dot{H}+4H^2+\frac{\omega}{3}F^2+2\frac{k}{a^2}=\frac{2}{3}BF
\end{equation}
and
\begin{equation}\label{11}
\dot{F}+F^2+3HF=-BF\, ,
\end{equation}
which are~not independent equations and where $H\equiv\dot{a}/a$, $B\equiv\dot{b}/b$ and
$F\equiv\dot{\phi}/\phi$. By employing relation (\ref{6.4}), one can
interpret the right hand side of equations (\ref{8}) and (\ref{9})
as energy density and pressure of the induced effective perfect
fluid, i.e.
\begin{equation}\label{11.1}
\rho_{_{\rm BD}}\equiv -T^{^{\rm (BD)}t}\
_{t}=-\frac{\phi}{8\pi}\left(3HB+BF\right)
\end{equation}
and
\begin{equation}\label{11.2}
p_{_{\rm BD}}\equiv T^{^{\rm (BD)}i}\
_{i}=\frac{\phi}{8\pi}(\dot{B}+B^2+2HB+BF)=-\frac{\phi}{8\pi}HB\,
,
\end{equation}
where $i=1$ or $2$ or $3$ without summation on it. The latter
equality in (\ref{11.2}) comes from equation (\ref{22}) which will
be derived in the next section. Therefor, the equation of state is
\begin{equation}\label{11.6}
p_{_{\rm BD}}=w_{_{\rm eff}}\rho_{_{\rm BD}} \qquad {\rm with}
\qquad w_{_{\rm eff}}=\frac{1}{F/H+3}\, .
\end{equation}

The usual matter in our universe has a positive energy density,
this basically has been demanded by the WEC, in which time--like
observers must obtain positive energy densities. Actually, the
complete WEC is~\cite{bookHowking1973}
\begin{equation}\label{11.3}
\left\{
  \begin{array}{ll}
    \rho_{_{\rm BD}}\geq0 \\
   \rho_{_{\rm BD}}+p_{_{\rm BD}}\geq0\, .
  \end{array}
\right.
\end{equation}
Now, let us consider that the scale factor of the fifth dimension and the
scalar field are~not constant values, i.e. $B\neq0$ and
$F\neq0$. Then, by applying conditions (\ref{11.3}) into relations
(\ref{11.1}) and (\ref{11.2}), one gets
\begin{equation}\label{11.4}
\left\{
  \begin{array}{ll}
    B>0 \\
    F\leq-4H
  \end{array}
\right.
\end{equation}
or
\begin{equation}\label{11.5}
\left\{
  \begin{array}{ll}
    B<0  \\
    F\geq-3H\, ,
  \end{array}
\right.
\end{equation}
where we also have assumed expanding universes, i.e. $H>0$.
Using conditions (\ref{11.4}) and (\ref{11.5}) in relation (\ref{11.6}) gives
\begin{equation}\label{11.8}
-1\leq w_{_{\rm eff}}\leq0
\end{equation}
or
\begin{equation}\label{11.9}
w_{_{\rm eff}}\geq0\, ,
\end{equation}
in where the effective dust matter can be achieved when $F/H$ goes to
negative or positive infinity, respectively.

In Section~$5$, we explore characteristic of the corresponding universes for the
above results. Meanwhile, in the following, we consider two special cases of a constant
scale factor of the fifth dimension and a constant scalar field.
\\ \\
\textbf{Constant Scale Factor of Fifth Dimension}
\\

When $b$ is a constant, equations (\ref{8})--(\ref{11}) reduce to
\begin{equation}\label{12}
H^2-\frac{\omega}{6}F^2+HF+\frac{k}{a^2}=0\, , \qquad 2\dot{H}+4H^2+\frac{\omega}{3}F^2+2\frac{k}{a^2}=0
 \qquad {\rm and} \qquad \dot{F}+F^2+3HF=0\, .
\end{equation}
These are exactly the ordinary vacuum BD equations in $4D$ space--time,
with $\rho_{_{\rm BD}}=0=p_{_{\rm BD}}$, as expected.
\\ \\
\textbf{Constant Scalar Field}
\\

When $\phi$ is a constant, action (\ref{1}) reduces to a $5D$
Einstein gravitational theory that has been considered in
Ref.~\cite{WessPonce92} in general situation (i.e. the extra
dimension is~not cyclic). In this case, equations
(\ref{8})--(\ref{11}) become
\begin{equation}\label{15}
H^2+\frac{k}{a^2}=-HB\,, \qquad \dot{H}+2H^2+\frac{k}{a^2}=0 \qquad {\rm and} \qquad \dot{B}+B^2+3HB=0\, .
\end{equation}
And, the usual FRW equations are equipped with $p_{_{\rm
BD}}=\rho_{_{\rm BD}}/3\equiv-HB/8\pi G$, which refers to a
radiation--like dominated universe for any kind of geometry
without a \emph{priori} assumption that the scale factor of the
fifth dimension is proportional to the inverse of the usual scale
factor, i.e. $b\propto a^{-1}$. Actually, the radiation--like
result is expected. For where there is no dependency on the extra
dimension, the usual four dimensional part of metric (\ref{7}) and
the third equation (\ref{15}) give a wave equation for the scale
factor of fifth dimension. Hence, definitions (\ref{6.5}) and
(\ref{6.6}) yield a traceless induced energy--momentum tensor, as
mentioned in Ref.\cite{WessPonce92}.

Exact solution of the second equation of (\ref{15}) is
\begin{equation}\label{19}
a=\sqrt{-kt^2+\alpha t}\, .
\end{equation}
Substituting solution (\ref{19}) into the first or third equation
of (\ref{15}) gives
\begin{equation}\label{21}
b=|\beta\dot{a}|=\left|\beta\frac{-2kt+\alpha}{2\sqrt{-kt^{2}+\alpha t}}
\right|=\left|\beta\frac{\sqrt{\alpha^{2}-4ka^{2}}}{2a}\right|\, ,
\end{equation}
where $\alpha$ and $\beta$ are constants of integration, and we
have assumed that $4D$ space--time has originated from a big bang.

For a closed geometry, solution (\ref{19}) admits $\alpha>0$ and
predicts a big crunch at $t=\alpha$ for the usual spatial
coordinates while the fifth dimension tends to infinite size and
is always real, for the maximum value of the usual scale factor is
$\alpha/2$. But, a flat geometry expands for ever and accepts
$\alpha>0$. An open geometry also expands for ever and admits
$\alpha\geq0$. In this case, $\alpha=0$ results in $a=t$ and
$b=|\beta|$. Time evolution of scale factors correspond to closed,
flat and open geometries have been illustrated in Fig.~$0$ with
constant values of $\alpha=1$ and $\beta=1$ as an example.

In the next two sections, we again consider a more general
situation in which the scale factor of the fifth dimension and the
scalar field are~not constants.
%%%%%%%%%%%%%%%%%%%%%%%%%%%%%%%%%%%%%%%%%%%%%%%%%%%%%%%%%%%%%%00000000
\begin{figure}
\begin{center}
\begin{tabular}{ccc}
\epsfig{figure=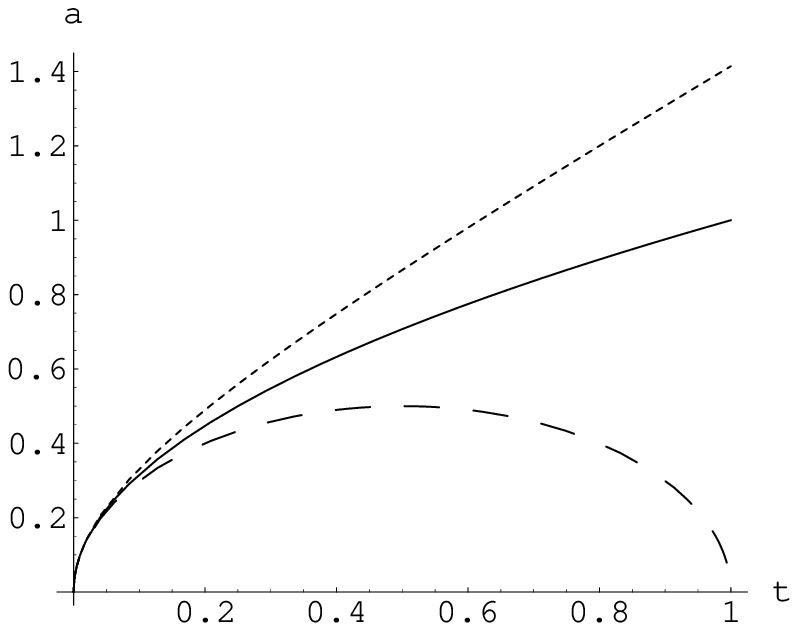,width=6cm}\footnotesize(i)
 \qquad
\epsfig{figure=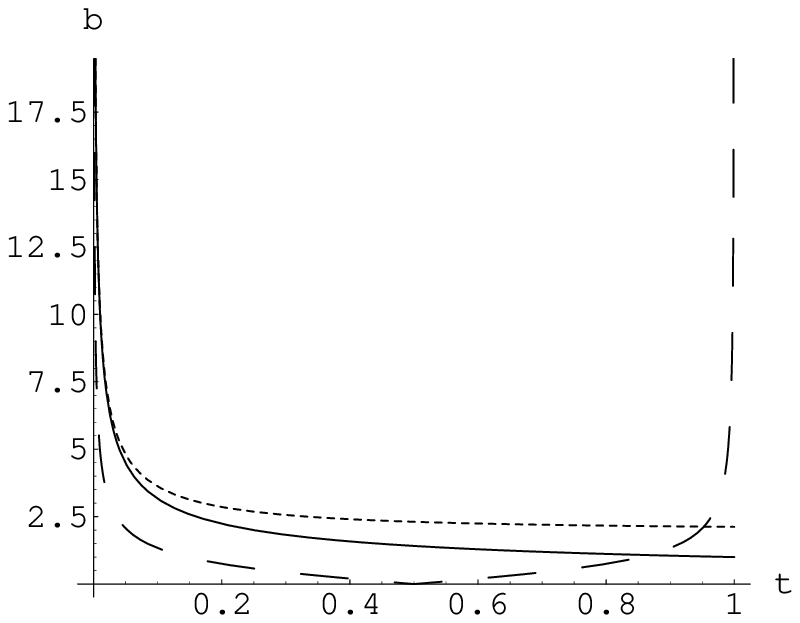,width=6cm}\footnotesize(ii)\vspace{1cm}
\end{tabular}

\hspace{0.5cm}{\footnotesize Figure 0: Time evolution of
scale factors (i) $a$ and (ii) $b$ for the special case of a
constant scalar field. The dashed, solid and dotted lines
correspond to closed, flat and open universes,
respectively.{}\qquad{}\quad{} }
\end{center}
\end{figure}
%%%%%%%%%%%%%%%%%%%%%%%%%%%%%%%%%%%%%%%%%%%%%%%%%%%%%%%%%%%%%%%

\section{Correspondence Between $5D$ Equations and $4D$ Ones}
\indent

Let us explore an equation (if any) similar to equation (\ref{11})
-- which is an integrable equation -- for when the rules of $F$
and $B$ are replaced. For this purpose, adding equations (\ref{8}) and
(\ref{9}), then subtracting equations (\ref{10}) and (\ref{11}) from it,
yields
\begin{equation}\label{22}
\dot{B}+B^2+3HB+BF=0.
\end{equation}
Comparing equations (\ref{11}) and (\ref{22}) shows that they are
equivalent to each other if one replaces $B$ by $F$. Indeed,
integrating equations (\ref{11}) and (\ref{22}) gives
\begin{equation}\label{23}
\dot{\phi}a^{3}b=m_{1} \qquad {\rm and} \qquad \dot{b}a^{3}\phi=m_{2}\, ,
\end{equation}
where $m_{1}\neq0$ and $m_{2}\neq0$ are constants of integration in general
situations when $\phi$ and $b$ are~not constants.
Actually, vanishing $m_{1}$ or $m_{2}$ gives $\phi$ or $b$ to be a
constant value, respectively, which have been discussed in the previous section.
Dividing equations (\ref{23}) by each
other leads to
\begin{equation}\label{25}
B=m'F\, ,
\end{equation}
where $m'\equiv m_{2}/m_{1}$. Relation (\ref{25}) obviously gives
\begin{equation}\label{26}
b=b_{o}\left(\frac{\phi}{\phi_{o}}\right)^{m'},
\end{equation}
where $b_{o}$ and $\phi_{o}$ are initial values.

Now, considering relation (\ref{25}), equations (\ref{8})--(\ref{11})
lead to three independent equations
\begin{equation}\label{27}
H^2-\frac{\tilde{\omega}}{6}\tilde{F}^2+H\tilde{F}+\frac{k}{a^2}=0\, ,
 \qquad 2\dot{H}+4H^2+\frac{\tilde{\omega}}{3}\tilde{F}^2+2\frac{k}{a^2}=0
  \qquad {\rm and} \qquad \dot{\tilde{F}}+\tilde{F}^2+3H\tilde{F}=0\, ,
\end{equation}
with
\begin{equation}\label{31}
\tilde{\phi} \equiv \phi^{m'+1} \qquad {\rm and} \qquad \tilde{\omega}\equiv\frac{\omega-2m'}{(m'+1)^2}\, ,
\end{equation}
where $m'\neq-1$ and $\tilde{F}=(m'+1)F$. Equations (\ref{27}) are
exactly the FRW equations of $4D$ vacuum BD theory. However, one
also needs to check if this new scalar field is a wave function in
$4D$ vacuum as well. For this purpose, it is easy to show that
\begin{equation}\label{31.1}
\Box \tilde{\phi}=(m'+1)
\left(\phi^{m'}_{,\mu}\phi^{,\mu}+\phi^{m'} \Box \phi \right)=0\,
,
\end{equation}
where equations (\ref{freduction}) and (\ref{7}) for a cyclic
extra dimension have been employed to get the second equality.

Hence, this procedure exhibits that $5D$ vacuum FRW--like
equations, equations (\ref{8})--(\ref{11}), are equivalent to the
corresponding $4D$ vacuum ones, equations (\ref{27}), with the
same spatial scale factor but a new (or modified) scalar field and
a new coupling constant, $\tilde{\phi}$ and $\tilde{\omega}$, in
which to have a non--ghost scalar field one must have
$\tilde{\omega}>-3/2$~\cite{Freund1982,16}.

For the special case of $m'=-1$, i.e. when $b\propto\phi^{-1}$,
equations (\ref{8})--(\ref{11}) reduce to
\begin{equation}\label{32}
H^2-\left(\frac{\omega}{6}+\frac{1}{3}\right)F^2+\frac{k}{a^2}=0\, ,
 \qquad 2\dot{H}+3H^2+\left(\frac{\omega}{2}+1\right)F^2+\frac{k}{a^2}=0
 \qquad {\rm and} \qquad \dot{F}+3HF=0\, .
\end{equation}
From the third equation of (\ref{32}) one gets
\begin{equation}\label{35}
F=F_{o}\left(\frac{a_{o}}{a}\right)^{3}.
\end{equation}
Using relation (\ref{35}) into the first and second equations of (\ref{32}) yields
\begin{equation}\label{36}
\ddot{a}a^{5}=2A\, \qquad {\rm and} \qquad (\dot{a}^{2}+k)a^{4}=-A,
\end{equation}
for $\dot{a}\neq0$ and $\omega\neq-2$ and where
$A\equiv-(\omega+2)F_{o}^{2}a_{o}^{6}/6$. These equations, or
actually their division i.e. $\ddot{a}a+2(\dot{a}^{2}+k)=0$, can
be solved by non--algebraic procedures, and their solutions
include the inverse--elliptic functions, although we do~not
perform it further. For a static universe, i.e. $\dot{a}=0$,
equations (\ref{32}) lead to a flat universe with $\omega=-2$. On
the other hand, if $\omega=-2$, then equations (\ref{32}) give
$\ddot{a}=0$ and $\dot{a}^{2}=-k$ which restrict the geometry
either to be flat or open. For $k=0$, one again gets a static
universe with $b=b_{o}\exp(-F_{o}t)$ and
$\phi=\phi_{o}\exp(F_{o}t)$. In the case $k=-1$, it leads to a
uniform expanding universe with $a=t$, and the evolution of scale
factor of the fifth dimension is
$b=b_{o}\exp(F_{o}a_{o}^{3}/2t^{2})$.

In the next section we continue our investigations for cosmological implications of equations
(\ref{8})--(\ref{11}) for a flat universe compatible with the recent observations.

\section{Exact Solutions for Flat Universe Compatible with Observations}
\indent

Measurements of anisotropies in the cosmic microwave background
radiation indicate that the universe is spatially flat~\cite{13},
so we concentrate on solutions with flat $3$--spaces. Therefor,
equations (\ref{8})--(\ref{10}) yield
\begin{equation}\label{37}
\dot{H}+3H^2+(B+F)H=0\, ,
\end{equation}
that gives
\begin{equation}\label{38}
\dot{a}a^{2}b\phi=m_{3}\, ,
\end{equation}
where $m_{3}$ is an integration constant. The case of vanishing
$m_{3}$ gives a static universe which is~not compatible with observations.
In general, relations (\ref{23}) and (\ref{38}) lead to
\begin{equation}\label{39}
b=b_{o}\left(\frac{a}{a_{o}}\right)^n \qquad {\rm and} \qquad \phi=\phi_{o}\left(\frac{a}{a_{o}}\right)^m,
\end{equation}
for $m_{3}\neq0$ and where $m\equiv m_{1}/m_{3}$ and $n\equiv m_{2}/m_{3}= mm'$,
also for general situations $B\neq0$ and $F\neq0$, we have $n\neq0$ and $m\neq0$.

Indeed, if in \emph{priori}, one had assumed $b\propto a^n$ (or $\phi\propto a^m$),
then equations (\ref{8})--(\ref{11}) would restrict the geometry to be spatially flat,
and automatically would give $\phi\propto a^m$ (or $b\propto a^n$).
Therefor, the power--law relation between the scale factor
of the fifth dimension and the scalar field with the usual scale
factor is a characteristic of the spatially flat universe.

Substituting solutions (\ref{39}) into equation (\ref{37}) gives
\begin{equation}\label{41}
\frac{\ddot{a}}{\dot{a}}+\left(m+n+2\right)\frac{\dot{a}}{a}=0.
\end{equation}
For $m+n\neq-3$, equation (\ref{41}) has a power--law solution
\begin{equation}\label{42}
a(t)=a_{o}\left(\frac{t}{t_{o}}\right)^{s} \qquad {\rm with} \qquad  H=\frac{s}{t}\, ,
\end{equation}
where $s\equiv(m+n+3)^{-1}$, and assumption expanding universes makes $s>0$.
Hence, solutions (\ref{39}) lead to
\begin{equation}\label{43}
b(t)=b_{o}\left(\frac{t}{t_{o}}\right)^{ns} \qquad {\rm with} \qquad B=\frac{ns}{t}
\end{equation}
and
\begin{equation}\label{44}
\phi(t)=\phi_{o}\left(\frac{t}{t_{o}}\right)^{ms} \qquad {\rm with} \qquad F=\frac{ms}{t}\, .
\end{equation}
There is also a constraint relation among the initial values,
namely $a_{o}^{3}b_{o}\phi_{o}=(m_{1}+m_{2}+3)t_{o}$.
Incidentally, the effective energy density and pressure, equations
(\ref{11.1}) and (\ref{11.2}) become
\begin{equation}\label{44.1}
\rho_{_{\rm BD}}=-\frac{\phi_{o}s^{2}}{8\pi
t_{o}^{ms}}n(m+3)t^{ms-2}
 \qquad {\rm and} \qquad p_{_{\rm BD}}=-\frac{\phi_{o}s^{2}}{8\pi t_{o}^{ms}}nt^{ms-2}.
\end{equation}

In the case $m+n=-3$, equations (\ref{39}) and (\ref{41}) give
exponential solutions
\begin{equation}\label{45}
a(t)=a_{o}e^{\lambda (t-t_{o})} \qquad {\rm with} \qquad H=\lambda\, ,
\end{equation}
\begin{equation}\label{46}
b(t)=b_{o}e^{n\lambda (t-t_{o})} \qquad {\rm with} \qquad B=n\lambda
\end{equation}
and
\begin{equation}\label{47}
\phi(t)=\phi_{o}e^{m\lambda (t-t_{o})} \qquad {\rm with} \qquad F=m\lambda\, ,
\end{equation}
where $\lambda$ is a constant and its positive values give
expanding universes, thus we assume $\lambda>0$. Incidentally, the
constraint relation among the initial values is
$a_{o}^{3}b_{o}\phi_{o}=m_{3}/\lambda$. In this case, the energy
density and pressure are
\begin{equation}\label{47.1}
\rho_{_{\rm BD}}=-\frac{\phi_{o}\lambda^{2}}{8\pi e^{m\lambda
t_{o}}}n(m+3)e^{m\lambda t} \qquad {\rm and} \qquad p_{_{\rm
BD}}=-\frac{\phi_{o}\lambda^{2}}{8\pi e^{m\lambda
t_{o}}}ne^{m\lambda t}.
\end{equation}
Note that, for both groups of solutions, the power law and
exponential ones, one has $w_{_{\rm eff}}=1/(3+m)$. We should emphasis that
all solutions of this section have been obtained
without a \emph{priori} ansatz for functionality of the scale factor and the scalar field.

In the next two subsections, we discuss properties of these
solutions. We should also remind that our vanishing induced
potential case is~not consistent with zero potential case of
Ref.~\cite{Ponce1and2} (where there, it requires $\omega=-1$
only).

\subsection{Power--Law Solutions}
\indent

Solutions are generally confined within some constraints that are
originated from mathematical or physical reasons. First of all, due to equations
(\ref{8})--(\ref{10}), the parameters $n$ and $m$
are~not independent. Substituting solutions (\ref{42})--(\ref{44})
into either of equations (\ref{8})--(\ref{10}) gives
\begin{equation}\label{48}
m_{\pm}=\frac{n+3\pm\sqrt{(n+3)^2+6\omega(n+1)}}{\omega}
\end{equation}
and hence
\begin{equation}\label{49}
s_{\pm}=\frac{\omega}{(\omega+1)(n+3)\pm\sqrt{(n+3)^2+6\omega(n+1)}}\, .
\end{equation}

Besides, our constraints are as follows. We have assumed $s>0$,
$m_{\pm}\neq0$, $n\neq0$ and $m_{\pm}+n\neq-3$ for power--law solutions. Real solutions of
relation (\ref{48}) dictate that $(n+3)^{2}+6\omega(n+1)\geq0$. By
substituting solutions (\ref{42})--(\ref{44}) in the WEC (\ref{11.4})
or (\ref{11.5}), we get
\begin{equation}\label{49.1}
\left\{
  \begin{array}{ll}
    n>0 \\
    m\leq-4
  \end{array}
\right.
\end{equation}
or
\begin{equation}\label{49.2}
\left\{
  \begin{array}{ll}
    n<0  \\
    m\geq-3\, ,
  \end{array}
\right.
\end{equation}
respectively. Note that, conditions (\ref{49.1}) and (\ref{49.2}) are compatible with
conditions (\ref{11.8}) and (\ref{11.9}), as expected.

In the following, we employ these constraints for when they lead
to cases of decelerated and especially accelerated universes.
Meanwhile, we should also remind that the deceleration parameter,
$q=-\ddot{a}a/\dot{a}^{2}$, in our model
for the power--law solutions is $q=1-s$.
\begin{description}
\item[] \textbf{Case Ia: Decelerated Universe}

It is supposed that the universe for a long time, when it was in
the radiation or dust dominated phases, was in a decelerating
regime. In our model, decelerating solutions can be obtained when
$0<s_{\pm}<1$. Acceptable domains of $n$ and $\omega$ for such a
range, without considering the WEC, is given in Table~$1$ and
Fig.~$1$. Note that, in Fig.~$1$, the part (ii) completely covers the part (i).
Also, adapted values of $n$ and $\omega$ with the WECs (\ref{49.1}) and
(\ref{49.2}) are shown in Table~$2$ with Fig.~$2$ and Table~$3$
with Fig.~$3$.
%%%%%%%%%%%%%%%%%%%%%%%%%%%%%%%%%%%%%%%%%%%%%%%%%%%%%%%%%%%%%%%%%%11111
\begin{table}
\begin{center}
\begin{tabular}{|c|c|c|}
  \hline
  $n$ & $\omega$ for $s_{+}$ values & $\omega$ for $s_{-}$ values \\ \hline\hline
  $n\leq-3$ & $\rm No \ solution$ & $-\frac{2(n^2+2n+3)}{(n+2)^2}<\omega<0$ \\
  $-3<n\leq-2$ & $0<\omega\leq-\frac{(n+3)^2}{6(n+1)}$ & $-\frac{2(n^2+2n+3)}
  {(n+2)^2}<\omega<0 \ \rm or\ 0<\omega\leq-\frac{(n+3)^2}{6(n+1)}$ \\
  $-2<n<-1$ & $\omega<-\frac{2(n^2+2n+3)}{(n+2)^2} \ \rm or\ 0<\omega\leq-\frac{(n+3)^2}
  {6(n+1)}$ & $\omega<0 \ \rm or\ 0<\omega\leq-\frac{(n+3)^2}{6(n+1)}$ \\
  $n=-1$ & $\omega<-4 \ \rm or\ \omega>0$ & $\rm No \ solution$ \\
  $-1<n<0$ & $-\frac{(n+3)^2}{6(n+1)}\leq\omega<-\frac{2(n^2+2n+3)}
  {(n+2)^2} \ \rm or\ \omega>0$ & $-\frac{(n+3)^2}{6(n+1)}\leq\omega<0 \ \rm or\ \omega>0$ \\
  $0<n\leq1$ & $\omega>0$ & $-\frac{2(n^2+2n+3)}{(n+2)^2}<\omega<0 \ \rm or\ \omega>0$ \\
  $n>1$ & $-\frac{(n+3)^2}{6(n+1)}\leq\omega<-\frac{2(n^2+2n+3)}{(n+2)^2}
  \ \rm or\ \omega>0$ & $-\frac{(n+3)^2}{6(n+1)}\leq\omega<0 \ \rm or\ \omega>0$ \\
  \hline
\end{tabular}
\caption{{\footnotesize Ranges of $n$ and $\omega$ for decelerating power--law solutions.}}
\end{center}
\end{table}
\begin{figure}
\begin{center}
\epsfig{figure=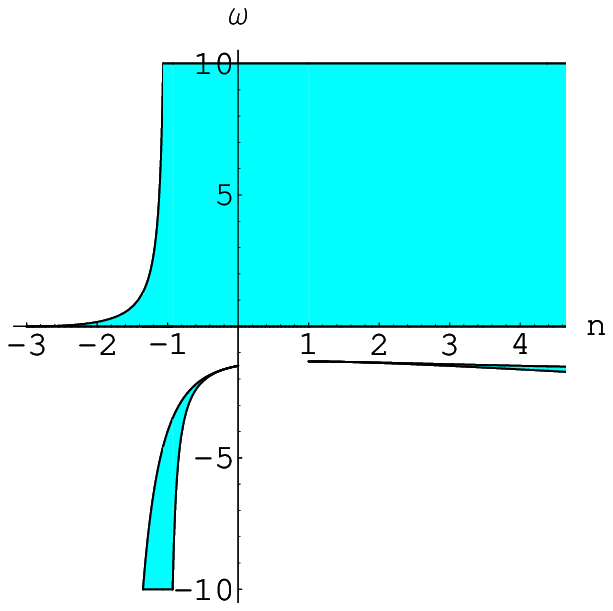,width=5.3cm}\footnotesize(i) \qquad
\epsfig{figure=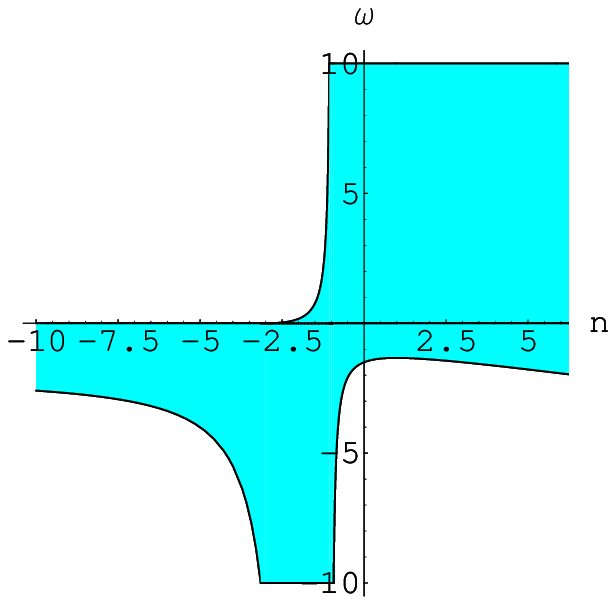,width=5.3cm}\footnotesize(ii)
\caption{{\footnotesize Domains of $n$ and $\omega$ correspond to
Table~$1$, (i) $s_{+}$ and (ii) $s_{-}$ values. Note that, the
line $n=-1$ is excluded in part (ii).}}
\end{center}
\end{figure}
%%%%%%%%%%%%%%%%%%%%%%%%%%%%%%%%%%%%%%%%%%%%%%%%%%%%%%%%%%%%%%%%%%%%22222
\begin{table}
\begin{center}
\begin{tabular}{|c|c|c|}
  \hline
  $n$ & $\omega$ for $s_{+}$ values & $\omega$ for $s_{-}$ values \\ \hline\hline
$2<n<3$ & $-\frac{n+9}{8}\leq\omega<-\frac{2(n^2+2n+3)}{(n+2)^2}$ & $\rm No \ solution$ \\
$n\geq3$ &
$-\frac{(n+3)^2}{6(n+1)}\leq\omega<-\frac{2(n^2+2n+3)}{(n+2)^2}$
& $-\frac{(n+3)^2}{6(n+1)}\leq\omega\leq-\frac{n+9}{8}$ \\ \hline
\end{tabular}
\caption{{\footnotesize Ranges of $n$ and $\omega$ for
decelerating power--law solutions which adapt the WEC
(\ref{49.1}).}}
\end{center}
\end{table}
\begin{figure}
\begin{center}
\epsfig{figure=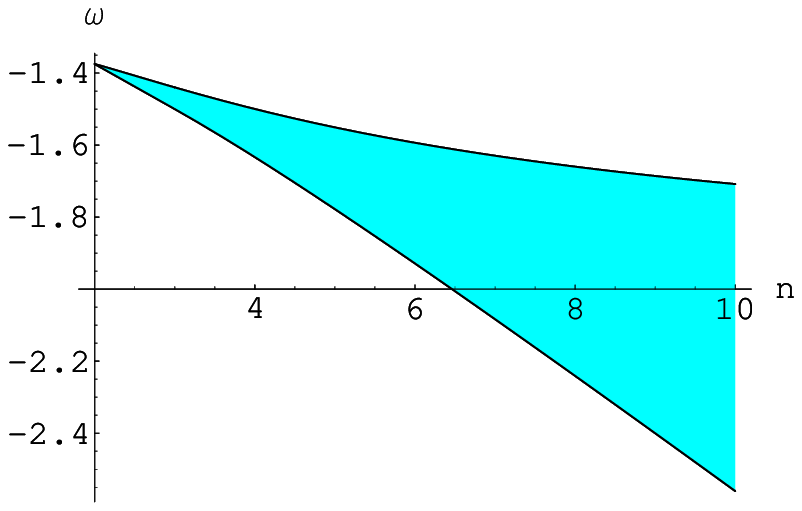,width=5.9cm}\footnotesize(i) \qquad
\epsfig{figure=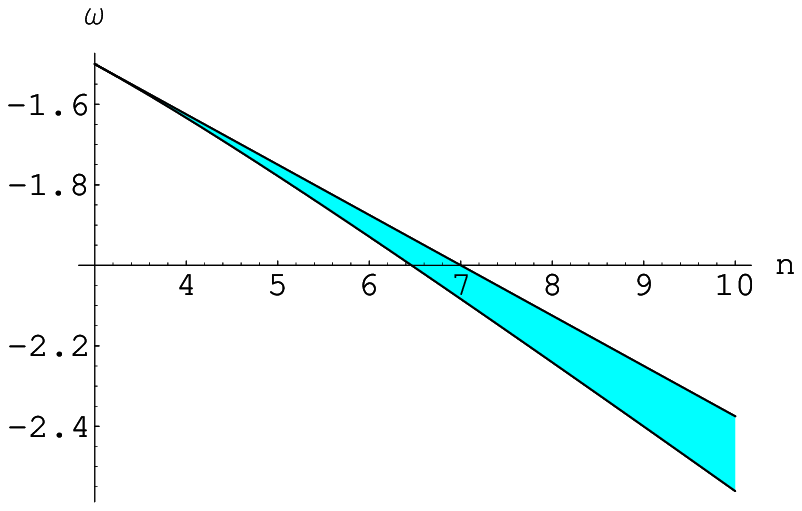,width=5.9cm}\footnotesize(ii)
\caption{{\footnotesize Domains of $n$ and $\omega$ correspond to
Table~$2$, (i) $s_{+}$ and (ii) $s_{-}$ values.}}
\end{center}
\end{figure}
%%%%%%%%%%%%%%%%%%%%%%%%%%%%%%%%%%%%%%%%%%%%%%%%%%%%%%%%%%%%%%%%3333333333
\begin{table}
\begin{center}
\begin{tabular}{|c|c|c|}
  \hline
  $n$ & $\omega$ for $s_{+}$ values & $\omega$ for $s_{-}$ values \\ \hline\hline
  $n\leq-3$ & $\rm No \ solution$ & $-\frac{2(n^2+2n+3)}{(n+2)^2}<\omega<0$ \\
  $-3<n\leq-2$ & $0<\omega\leq-\frac{(n+3)^2}{6(n+1)}$ & $-\frac{2(n^2+2n+3)}
  {(n+2)^2}<\omega<0 \ \rm or \ 0<\omega\leq-\frac{(n+3)^2}{6(n+1)}$ \\
  $-2<n<-1$ & $\omega<-\frac{2(n^2+2n+3)}{(n+2)^2} \ \rm or\ 0<\omega\leq-\frac{(n+3)^2}
  {6(n+1)}$ & $\omega<0\ \rm or\ 0<\omega\leq-\frac{(n+3)^2}{6(n+1)}$ \\
  $n=-1$ & $\omega<-4 \ \rm or\ \omega>0$ & $\rm No \ solution$ \\
  $-1<n<0$ & $-\frac{(n+3)^2}{6(n+1)}\leq\omega<-\frac{2(n^2+2n+3)}{(n+2)^2}
  \ \rm or\ \omega>0$ & $-\frac{(n+3)^2}{6(n+1)}\leq\omega<0\ \rm or\ \omega>0$ \\  \hline
\end{tabular}
\end{center}
\caption{\footnotesize Ranges of $n$ and $\omega$ for decelerating
power--law solutions which adapt the WEC (\ref{49.2}).}
\end{table}
\begin{figure}
\begin{center}
\epsfig{figure=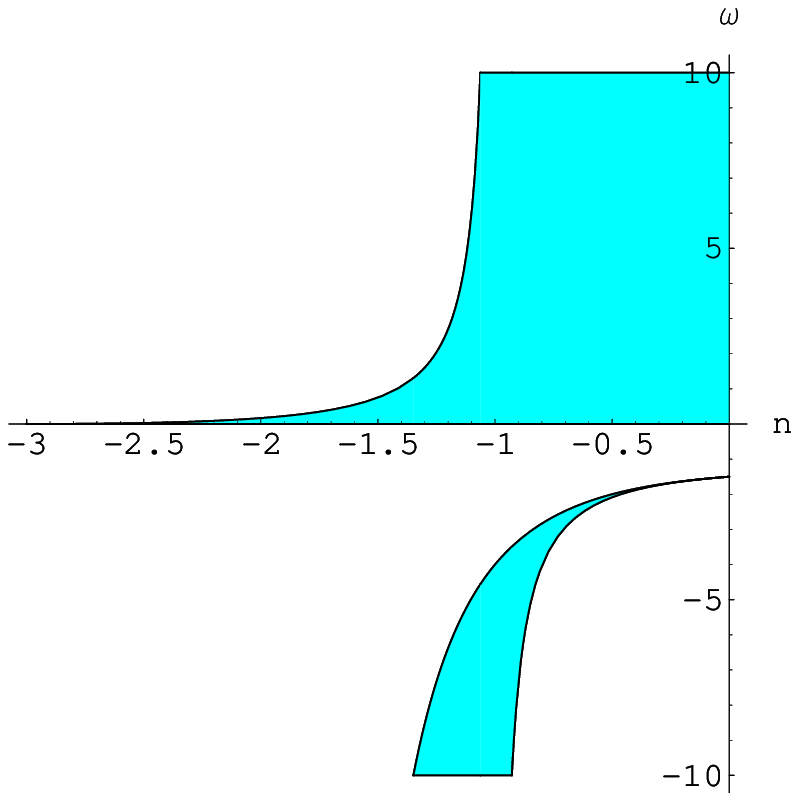,width=4.6cm}\footnotesize(i) \qquad
\epsfig{figure=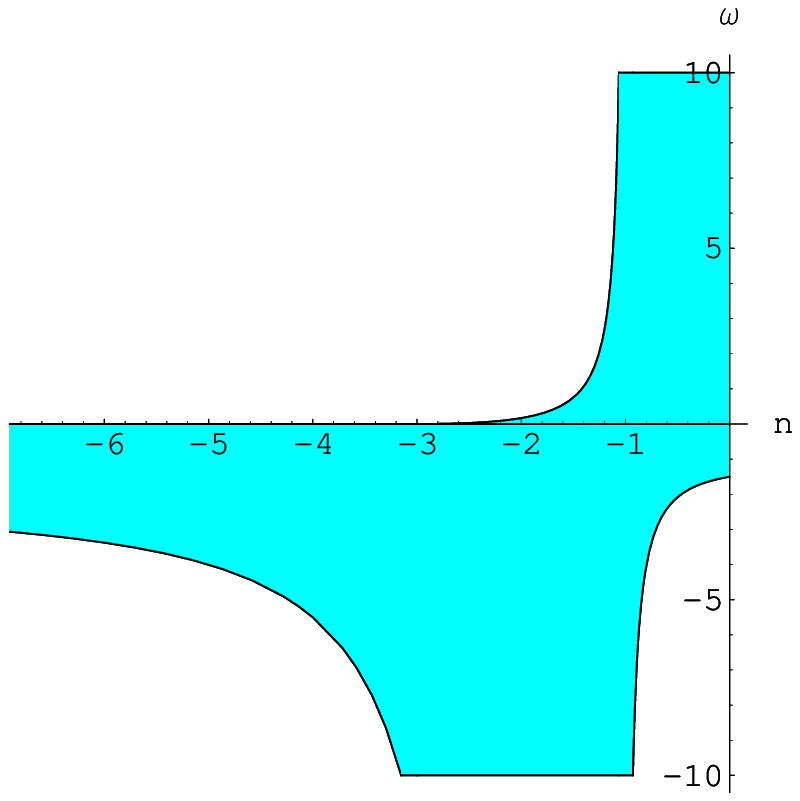,width=4.6cm}\footnotesize(ii)
\caption{\footnotesize Domains of $n$ and $\omega$ correspond to
Table~$3$, (i) $s_{+}$ and (ii) $s_{-}$ values. Note that, the
line $n=-1$ is excluded in part (ii).}
\end{center}
\end{figure}
%%%%%%%%%%%%%%%%%%%%%%%%%%%%%%%%%%%%%%%%%%%%%%%%%%%%%%%%%%%%%%%%%%%%%%%
\newpage
\item[] \textbf{Case IIa: Accelerated Universe}

Recent observations show that the universe is in an accelerating
regime at the present epoch~\cite{14}. This makes $s_{\pm}>1$, and acceptable values of $n$
and $\omega$ corresponding to this condition, without considering the
WEC, are given in Table~$4$ and Fig.~$4$. In Fig.~$4$, the maximum
value of $\omega$ for $s_{+}$ values tends to $-5/4$ when $n=1$, and for
$s_{-}$ values tends to $-4/3$ when $n\rightarrow1^{-}$. Corresponding cases with the
WECs (\ref{49.1}) and (\ref{49.2}) are illustrated in Table~$5$
with Fig.~$5$ and Table~$6$ with Fig.~$6$, respectively.
\end{description}
%%%%%%%%%%%%%%%%%%%%%%%%%%%%%%%%%%%%%%%%%%%%%%%%%%%%%%%%%%%%%%%%%%%%%%%%%44444444444
\begin{table}
\begin{center}
\begin{tabular}{|c|c|c|}
  \hline
  $n$ & $\omega$ for $s_{+}$ values & $\omega$ for $s_{-}$ values \\ \hline\hline
  $n\leq-3$ & $\rm No \ solution$ & $-\frac{2(n^2+3n+6)}{(n+3)^2}<\omega<-\frac{2(n^2+2n+3)}{(n+2)^2}$ \\
  $-3<n<-2$ & $\omega<-\frac{2(n^2+3n+6)}{(n+3)^2}$ & $\omega<-\frac{2(n^2+2n+3)}{(n+2)^2}$ \\
  $n=-2$ & $\omega<-8$ & $\rm No \ solution$ \\
  $-2<n<0$ & $-\frac{2(n^2+2n+3)}{(n+2)^2}<\omega<-\frac{2(n^2+3n+6)}{(n+3)^2}$ & $\rm No \ solution$ \\
  $0<n<1$ & $-\frac{(n+3)^2}{6(n+1)}\leq\omega<-\frac{2(n^2+3n+6)}{(n+3)^2}$
          & $-\frac{(n+3)^2}{6(n+1)}\leq\omega<-\frac{2(n^2+2n+3)}{(n+2)^2}$ \\
  $n\geq1$ & $-\frac{2(n^2+2n+3)}{(n+2)^2}<\omega<-\frac{2(n^2+3n+6)}{(n+3)^2}$ & $\rm No \ solution$ \\
  \hline
\end{tabular}
\end{center}
\caption{\footnotesize Ranges of $n$ and $\omega$ for accelerating power--law solutions.}
\end{table}
\begin{figure}
\begin{center}
\epsfig{figure=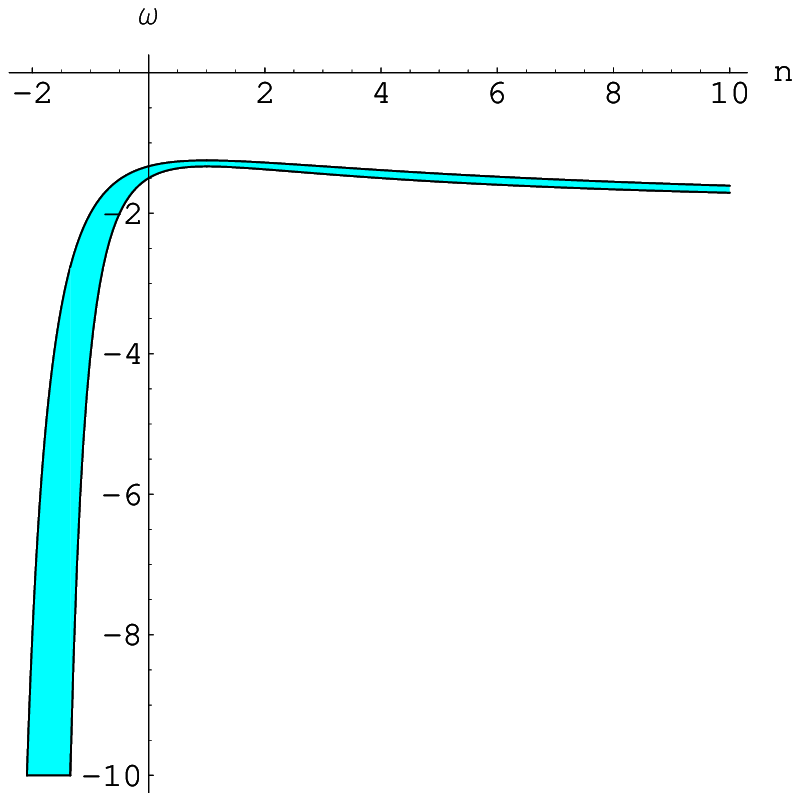,width=5cm}\footnotesize(i) \qquad
\epsfig{figure=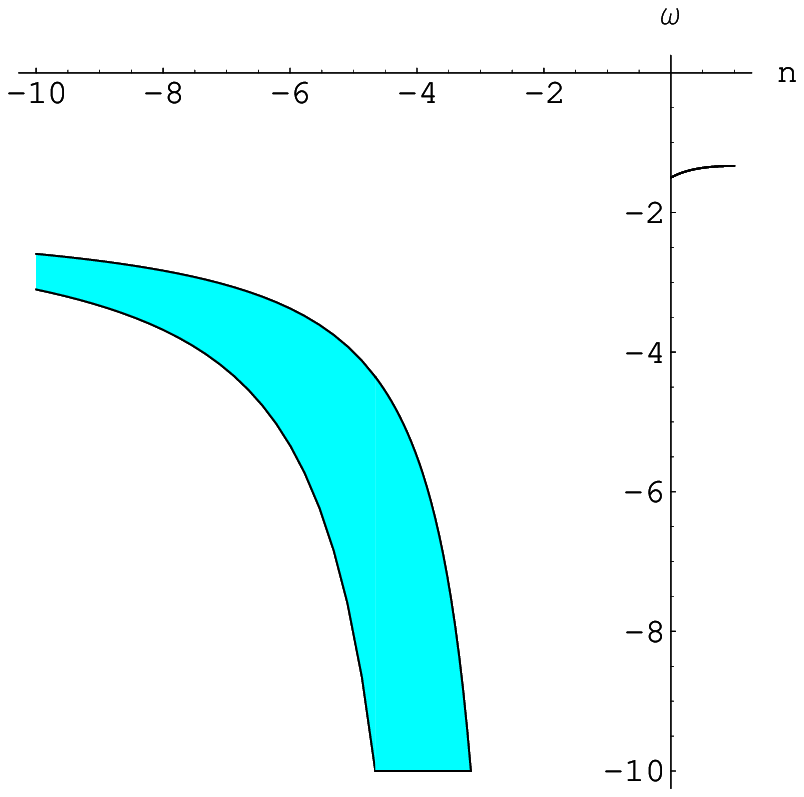,width=5cm}\footnotesize(ii)
\caption{\footnotesize Domains of $n$ and $\omega$ correspond to
Table~$4$, (i) $s_{+}$ and (ii) $s_{-}$ values.}
\end{center}
\end{figure}
%%%%%%%%%%%%%%%%%%%%%%%%%%%%%%%%%%%%%%%%%%%%%%%%%%%%%%%%%%%%%%%%%%%%%%%%%555555555555
\begin{table}
\begin{center}
\begin{tabular}{|c|c|c|}
  \hline
  $n$ & $\omega$ for $s_{+}$ values & $\omega$ for $s_{-}$ values \\ \hline\hline
$1<n<2$ & $-\frac{n+9}{8}\leq\omega<-\frac{2(n^2+3n+6)}{(n+3)^2}$ & $\rm No \ solution$ \\
$n\geq2$ &
$-\frac{2(n^2+2n+3)}{(n+2)^2}<\omega<-\frac{2(n^2+3n+6)}{(n+3)^2}$
& $\rm No \ solution$ \\ \hline
\end{tabular}
\end{center}
\caption{\footnotesize Ranges of $n$ and $\omega$ for accelerating
power--law solutions which adapt the WEC (\ref{49.1}).}
\end{table}
\begin{figure}
\begin{center}
\epsfig{figure=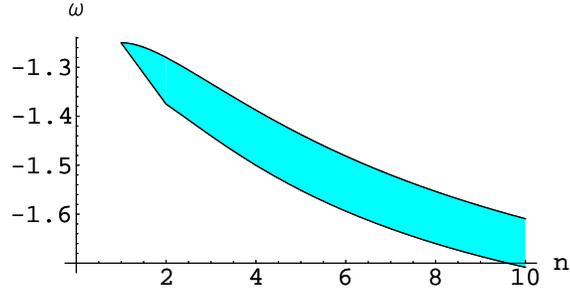,width=8cm}\footnotesize
\caption{\footnotesize Domains of $n$ and $\omega$ correspond to
Table~$5$ for $s_{+}$ values.}
\end{center}
\end{figure}
%%%%%%%%%%%%%%%%%%%%%%%%%%%%%%%%%%%%%%%%%%%%%%%%%%%%%%%%%%%%%%%%%%%%%%%%%666666666666
\begin{table}
\begin{center}
\begin{tabular}{|c|c|c|}
  \hline
  $n$ & $\omega$ for $s_{+}$ values & $\omega$ for $s_{-}$ values \\ \hline\hline
  $n\leq-3$ & $\rm No \ solution$ & $-\frac{2(n^2+3n+6)}{(n+3)^2}<\omega<-\frac{2(n^2+2n+3)}{(n+2)^2}$ \\
  $-3<n<-2$ & $\omega<-\frac{2(n^2+3n+6)}{(n+3)^2}$ & $\omega<-\frac{2(n^2+2n+3)}{(n+2)^2}$ \\
  $n=-2$ & $\omega<-8$ & $\rm No \ solution$ \\
  $-2<n<0$ & $-\frac{2(n^2+2n+3)}{(n+2)^2}<\omega<-\frac{2(n^2+3n+6)}{(n+3)^2}$ & $\rm No \ solution$ \\
  \hline
\end{tabular}
\end{center}
\caption{\footnotesize Ranges of $n$ and $\omega$ for accelerating
power--law solutions which adapt the WEC (\ref{49.2}).}
\end{table}
\begin{figure}
\begin{center}
\epsfig{figure=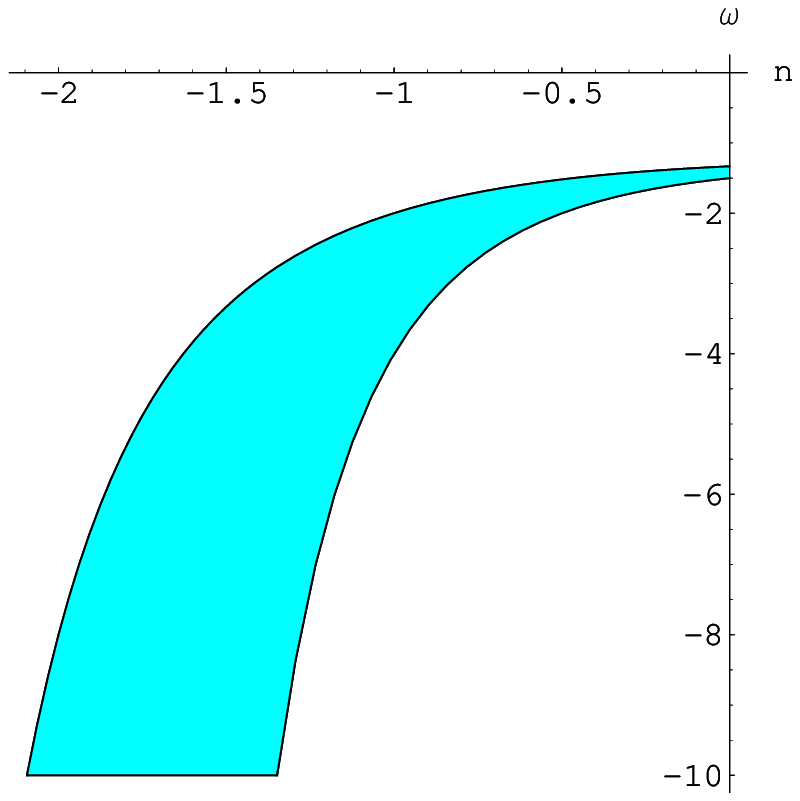,width=4.7cm}\footnotesize(i) \qquad
\epsfig{figure=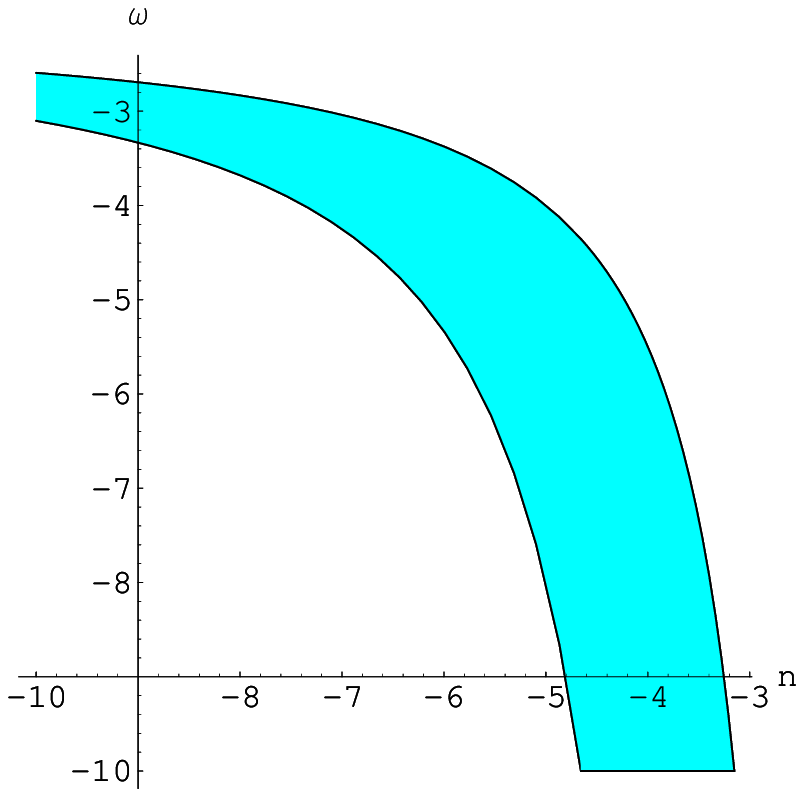,width=4.7cm}\footnotesize(ii)
\caption{\footnotesize Domains of $n$ and $\omega$ correspond to
Table~$6$, (i) $s_{+}$ and (ii) $s_{-}$ values.}
\end{center}
\end{figure}
%%%%%%%%%%%%%%%%%%%%%%%%%%%%%%%%%%%%%%%%%%%%%%%%%%%%%%%%%%%%

It should be emphasized that though astronomical tests in the solar
system requires a positive large value for $\omega$, but still in the
large cosmological scale, one cannot definitely
rule out small or even negative values of the BD coupling constant.
Indeed, these values of $\omega$ have achieved considerable interests in the literature.

By considering the WEC (\ref{49.1}) in relations (\ref{44.1}),
one gets positive energy densities, as expected, but with negative pressures,
where both $\rho_{_{\rm BD}}$ and $|p_{_{\rm BD}}|$ decrease with the time.
In this case, even though the pressure is negative, but Figs.~$2$ and $5$ illustrate that
one has decelerating and accelerating solutions.
On the other hand, using condition (\ref{49.2}) into relations (\ref{44.1})
gives positive energy densities and pressures.
Although, in this situation the pressure is positive, but still Figs.~$3$ and $6$
indicate that one again has decelerating and accelerating solutions.
In this situation, for decreasing energy density and pressure
with the time, one has to restrict $ms<2$, which most of the solutions
fulfill it.

Yet we have one more condition, namely non--ghost scalar fields with
$\omega> -4/3$, to be imposed. With this situation, acceptable solutions
are as follows.
\begin{description}
\item[] \textbf{Case Ib: Decelerated Universe}

Acceptable values of $n$ and $\omega$ for the range
$\omega> -4/3$ restrict Table~$3$ and Fig.~$3$,
and the results are shown in Table~$7$ and Fig.~$7$.
Hence, this model admits a typical decelerated universe with non--ghost
scalar fields, positive induced energy density and pressure,
fulfilling the WEC (\ref{49.2}), where the scale factor of
fifth dimension shrinks with the time.
Incidentally, Fig.~$2$ illustrates that there is~not any decelerated solution with
non--ghost scalar fields which complies with the WEC (\ref{49.1}).

\item[] \textbf{Case IIb: Accelerated Universe}

Table~$8$ and Fig.~$8$, which are the reductions of Table~$5$ and Fig.~$5$,
illustrate the corresponding domains of $n$
and $\omega$ for $\omega> -4/3$. Therefore,
the model also admits a typical accelerated universe with
non--ghost scalar fields, positive induced energy density and
negative pressure, fulfilling the WEC (\ref{49.1}),
where the scale factor of fifth dimension grows with the time.
This situation restricts $1<n<3$, contrary to the assumption of $n=1$ in
Ref.~\cite{12}. Also, Fig.~$6$ indicates that accelerated solutions do~not exist for
non--ghost scalar fields which fulfill the WEC (\ref{49.2}).
\end{description}
%%%%%%%%%%%%%%%%%%%%%%%%%%%%%%%%%%%%%%%%%%%%%%%%%%%%%%%%%%%%%%%77777777
\begin{table}
\begin{center}
\begin{tabular}{|c|c|c|}
  \hline
  $n$ & $\omega$ for $s_{+}$ values & $\omega$ for $s_{-}$ values \\ \hline\hline
  $n\leq-3$ & $\rm No \ solution$ & $-\frac{4}{3}<\omega<0$ \\
  $-3<n<-1$ & $0<\omega\leq-\frac{(n+3)^2}{6(n+1)}$ & $-\frac{4}{3}<\omega<0
  \ \rm or\ 0<\omega\leq-\frac{(n+3)^2}{6(n+1)}$ \\
  $n=-1$ & $\omega>0$ & $\rm No \ solution$ \\
  $-1<n<0$ & $\omega>0$ & $-\frac{4}{3}<\omega<0 \ \rm or\ \omega>0$ \\ \hline
  \end{tabular}
\end{center}
\caption{\footnotesize Ranges of $n$ and $\omega$ for decelerating power--law solutions
with non--ghost scalar fields which adapt the WEC (\ref{49.2}). }
\end{table}
\begin{figure}
\begin{center}
\epsfig{figure=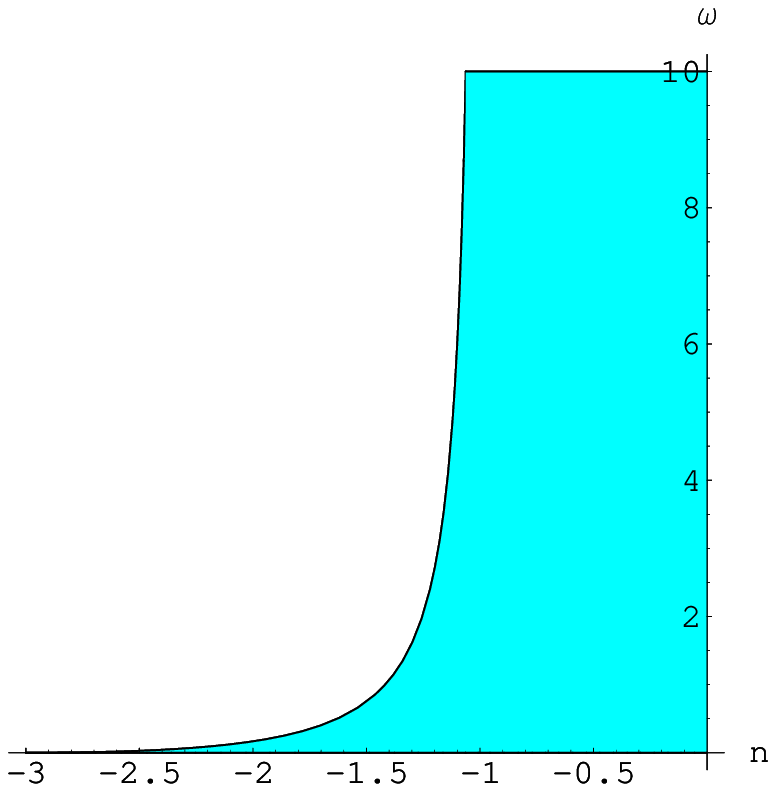,width=4.5cm}\footnotesize(i) \qquad
\epsfig{figure=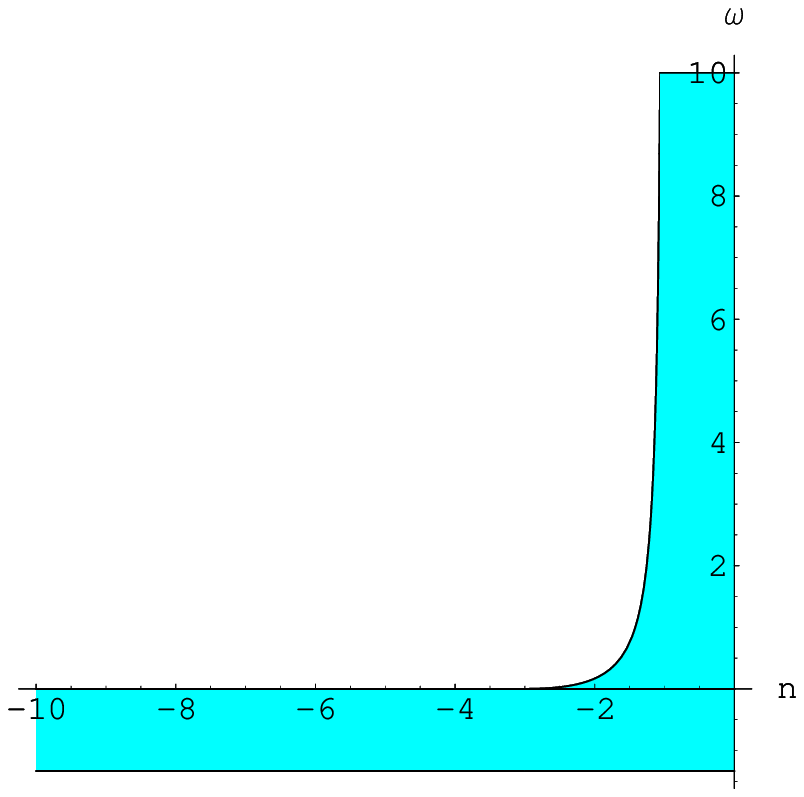,width=4.5cm}\footnotesize(ii)
\caption{\footnotesize Domains of $n$ and $\omega$ correspond to
Table~$7$, (i) $s_{+}$ and (ii) $s_{-}$ values. Note that, the
line $n=-1$ is excluded in part (ii).}
\end{center}
\end{figure}
%%%%%%%%%%%%%%%%%%%%%%%%%%%%%%%%%%%%%%%%%%%%%%%%%%%%%%%%%%%%%%%%%%%%%%88888888
\begin{table}
\begin{center}
\begin{tabular}{|c|c|c|}
  \hline
  $n$ & $\omega$ for $s_{+}$ values \\ \hline\hline
$1<n<\frac{5}{3}$ & $-\frac{n+9}{8}\leq\omega<-\frac{2(n^2+3n+6)}{(n+3)^2}$  \\
$\frac{5}{3}\leq n<3$ & $-\frac{4}{3}<\omega<-\frac{2(n^2+3n+6)}{(n+3)^2}$ \\ \hline
\end{tabular}
\end{center}
\caption{\footnotesize Ranges of $n$ and $\omega$ for
accelerating power--law solutions with non--ghost scalar fields which
fulfill the WEC (\ref{49.1}).}
\end{table}
\begin{figure}
\begin{center}
\epsfig{figure=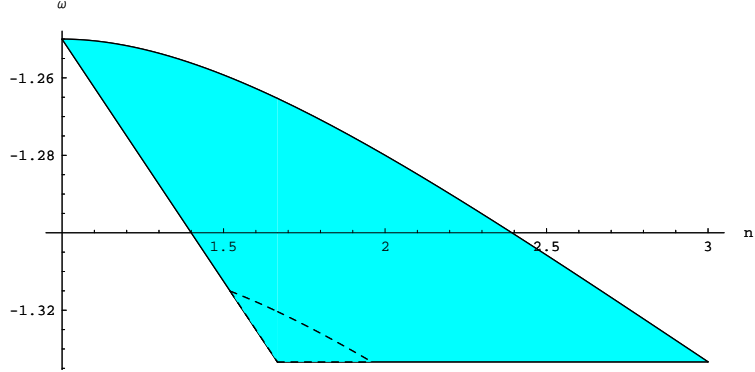,width=10cm}\footnotesize
\caption{\footnotesize Domains of $n$ and $\omega$ correspond to
Table~$8$. The area below the dashed line corresponds to
Table~$9$, and the upper border curve corresponds to Table~$10$
with non--ghost scalar fields.}
\end{center}
\end{figure}
%%%%%%%%%%%%%%%%%%%%%%%%%%%%%%%%%%%%%%%%%%%%%%%%%%%%%%%%%%%%%%%%%%%%%%%%%

By employing the recent observational measurements of $q$, namely $-0.92\leq q_{o}\leq
-0.42$~\cite{17}, we obtain $1.42\leq s \leq 1.92$. This range of
$s$ is only compatible with the accelerating Case~II, but
imposes more restrictions on domains of $n$ and $\omega$.
Indeed, Table~$8$ and Fig.~$8$ of Case~IIb with these values of $s$ lead to
Table~$9$ and the area below the dashed line in Fig.~$8$.
Thus, one obtains ranges $1.5208\leq n<1.9583$ and $-4/3<\omega\leq-1.3151$.
These are the best values of $n$ and $\omega$, in this model,
that are compatible with the recent observations.
These values also fulfill the condition $\tilde{\omega}>-3/2$, which is required,
in Section 4, for having non--ghost scalar fields in the equivalent $4D$
vacuum BD equations.
%%%%%%%%%%%%%%%%%%%%%%%%%%%%%%%%%%%%%%%%%%%%%%%%%%%%%%%%%%%%%%%%%%%%9999999
\begin{table}
\begin{center}
\begin{tabular}{|c|c|c|}
  \hline
  $n$ & $\omega$ for $s_{+}$ values \\ \hline\hline
$n=1.5208$ & $\omega=-1.3151$ \\
$1.5208\leq n<\frac{5}{3}$ & $-\frac{n+9}{8}\leq\omega\leq-\frac{32(144n^2+357n+639)}{2304n^2+11424n+14161}$ \\
$\frac{5}{3}\leq n<1.9583$ & $-\frac{4}{3}<\omega\leq-\frac{32(144n^2+357n+639)}{2304n^2+11424n+14161}
$ \\ \hline
\end{tabular}
\end{center}
\caption{\footnotesize Ranges of $n$ and $\omega$ for
accelerating power--law solutions with non--ghost scalar fields which
fulfill the WEC (\ref{49.1}) for $s_{+}$ values, and are compatible with the recent observations.}
\end{table}
%%%%%%%%%%%%%%%%%%%%%%%%%%%%%%%%%%%%%%%%%%%%%%%%%%%%%%%%%%%%%%%%%%%%%%%%%%%%%

It is interesting to note that, one can infer from relations
(\ref{48}) and (\ref{49}) that $m_{+}$ and $s_{+}$ do~not allow
$\omega=0$. However, the other constraints does~not obviously show
that the zero value of $\omega$ is prevented, but the mathematical
procedure for all tables and figures indicates that $\omega\neq0$.

\subsection{Exponential Solutions}
\indent

Relation (\ref{45}) represents an exponential growth of the scale
factor, i.e. an inflationary universe. However, no such a rapid expansion has been indicated at
present or throughout almost the whole history of the universe
except at the very early universe stage. Nevertheless, let us probe some
properties of these solutions.

First of all, we have $m+n=-3$ that, with assumptions $m\neq0$ and $n\neq0$, restricts
$m\neq-3$, $n\neq-3$ and $m\neq-n$.
By employing solutions (\ref{45})--(\ref{47}) into equations (\ref{8})--(\ref{10}), one
gets
\begin{equation}\label{50}
m_{\pm}=\frac{-3\pm\sqrt{-(12\omega+15)}}{\omega+2}\, .
\end{equation}
Substituting condition $m+n=-3$ into relation
(\ref{50}) gives
\begin{equation}\label{51}
\omega=-\frac{2(n^2+3n+6)}{(n+3)^2}\, ,
\end{equation}
for $n\neq0, -3$. Real solutions of relation (\ref{50}) impose
$\omega\leq-5/4$. Solutions (\ref{45})--(\ref{47}) satisfy
conditions (\ref{49.1}) and (\ref{49.2}) with more restrictions.
Using the WEC (\ref{49.1}), in relations (\ref{47.1}) gives
positive energy densities and negative pressures, where both
$\rho_{_{\rm BD}}$ and $|p_{_{\rm BD}}|$ decrease rapidly with the
time. But, considering condition (\ref{49.2}) in relations
(\ref{47.1}), gives positive energy densities and pressures. If
one takes $m<0$ in condition (\ref{49.2}), the energy density and
pressure again will decrease with the time.

Acceptable ranges of $n$ and $\omega$ for exponential solutions fulfilling the
WECs (\ref{49.1}) and (\ref{49.2}) are shown in Tables~$10$ and~$11$. These tables
indicate that, only for the range $1\leq n<3$, solutions do avoid ghost scalar
fields. In Fig.~$8$, the upper border curve illustrates the acceptable
values of $n$ and $\omega$ for an inflationary universe fulfilling
the WEC (\ref{49.1}) with non--ghost scalar fields.
%%%%%%%%%%%%%%%%%%%%%%%%%%%%%%%%%%%%%%%%%%%%%%%%%%%%%%%%%%%%%%%%10
\begin{table}
\begin{center}
\begin{tabular}{|c|c|c|}
  \hline
  $n$ & $\omega$ for $m_{+}$ values & $\omega$ for $m_{-}$ values \\ \hline\hline
$n=1$ & $\omega=-\frac{5}{4}$ & $\omega=-\frac{5}{4}$ \\
$n>1$ &
$\rm No \ solution$
& $\omega=-\frac{2(n^2+3n+6)}{(n+3)^2}$ \\ \hline
\end{tabular}
\end{center}
\caption{\footnotesize Ranges of $n$ and $\omega$ for exponential
solutions which fulfill the WEC (\ref{49.1}).}
\end{table}
%%%%%%%%%%%%%%%%%%%%%%%%%%%%%%%%%%%%%%%%%%%%%%%%%%%%%%%%%%%%%%%%%%11
\begin{table}
\begin{center}
\begin{tabular}{|c|c|c|}
  \hline
  $n$ & $\omega$ for $m_{+}$ values & $\omega$ for $m_{-}$ values \\ \hline\hline
$n<-3$ & $\rm No \ solution$ & $\omega=-\frac{2(n^2+3n+6)}{(n+3)^2}$ \\
$-3<n<-1$ & $\omega=-\frac{2(n^2+3n+6)}{(n+3)^2}$ & $\rm No \ solution$ \\
$-1<n<0$ & $\omega=-\frac{2(n^2+3n+6)}{(n+3)^2}$ & $\rm No \ solution$ \\ \hline
\end{tabular}
\end{center}
\caption{\footnotesize Ranges of $n$ and $\omega$ for exponential
solutions which fulfill the WEC (\ref{49.2}).}
\end{table}
%%%%%%%%%%%%%%%%%%%%%%%%%%%%%%%%%%%%%%%%%%%%%%%%%%%%%%%%%%%%%%%%%%%%%%%%
\section{Conclusions}
\indent

Analogous to the approach of IM theories, one can consider the BD
gravity as the underlying theory. Hence, extra geometrical terms,
coming from the fifth dimension, are regarded as an
induced--matter and induced potential. We have followed, with some
corrections, the procedure of Ref.~\cite{12} for introducing the
induced potential and have employed a generalized FRW type
solution for a $5D$ vacuum BD theory. Hence, the scalar field and
scale factors of the $5D$ metric can, in general, be functions of
the cosmic time and the extra dimension. However, for simplicity,
we have assumed the scalar field and scale factors to be only
functions of the cosmic time, where this makes the induced
potential, by its definition, vanishes.

We then have revealed that in general situations, in which the
scale factor of the fifth dimension and scalar field are~not
constants, the $5D$ equations, for any kind of geometry, admit a
power--law relation between the scalar field and scale factor of
the fifth dimension. Hence, the procedure exhibits that $5D$
vacuum FRW--like equations are equivalent, in general, to the
corresponding $4D$ vacuum ones with the same spatial scale factor
but a new (or modified) scalar field and a new coupling constant.
This equivalency can be viewed as the distinguished point of this
work from Refs.~\cite{12,Ponce1and2}. Indeed, through
investigating the $5D$ vacuum FRW--like equations, we have shown
that its equivalent $4D$ vacuum equations admit accelerated scale
factors, contrary to what one may have expected from a vacuum
space--time. Conclusions of the complete investigation of the
induced $4D$ equations are as follows.

Following our investigations for cosmological implications, we
have shown that for the special case of a constant scale factor of
the fifth dimension, the $5D$ vacuum FRW--like equations reduce to
the corresponding equations of the usual $4D$ vacuum BD theory, as
expected. In the special case of a constant scalar field, the
action reduces to a $5D$ Einstein gravitational theory and the
equations reduce to the usual FRW equations with a typical
radiation dominated universe. For this situation, we also have
obtained dynamics of scale factors of the ordinary and extra
dimensions for any kind of geometry without any \emph{priori}
assumption among them. Solutions predict a limited life time for
closed geometries and unlimited one for flat and open geometries.
A typical time evolutions of scale factors correspond to closed,
flat and open geometries have been illustrated in Fig.~$0$.

Then, we have focused on spatially flat geometries and have
obtained exact solutions of scale factors and scalar field.
Solutions are found to be in the form of power--law and
exponential ones in the cosmic time. We also have employed the WEC
for the induced--matter of the $4D$ modified BD gravity, that
gives two conditions (\ref{49.1}) and (\ref{49.2}). We then have
pursued properties of these solutions and have indicated
mathematically and physically acceptable ranges of them, and the
results have been presented in a few tables and figures.

All types of solutions fulfill the WECs in different ranges, where
the exponential solutions are more restricted. The solutions
fulfilling the WEC (\ref{49.1}) have negative pressures, but the
figures illustrate that for the power--law results there are
decelerating solutions beside accelerating ones. For this
condition, both $\rho_{_{\rm BD}}$ and $|p_{_{\rm BD}}|$ decrease
with the cosmic time, but the extra dimension grows. On the other
hand, the solutions satisfying the WEC (\ref{49.2}) have positive
pressures, where the power--law results accept accelerating
solutions in addition to decelerating ones. For this condition,
again decreasing energy density and pressure with the time can
occur for some solutions, however all with shrinking extra
dimension. The homogeneity between the extra dimension and the
usual spatial dimensions, i.e. $b\propto a$, can take place in the
solutions, but for the power--law ones the WECs exclude it.

By considering non--ghost scalar fields and appealing the recent
observational measurements, the solutions have been more
restricted. Actually, we have illustrated that the accelerating
power--law solutions, which satisfy the WEC and have non--ghost
scalar fields, are compatible with the recent observations in
ranges $-4/3<\omega\leq-1.3151$ for the BD coupling constant and
$1.5208\leq n<1.9583$ for dependence of the fifth dimension scale
factor with the usual scale factor. These ranges also fulfill the
condition $\tilde{\omega}>-3/2$ which prevents ghost scalar fields
in the equivalent $4D$ vacuum BD equations. Incidentally, this
range is more restricted than the one obtained in
Ref.~\cite{Ponce1and2}, i.e. $-1.5<\omega<-1$, where the
difference may have been caused by the distinct definition of the
induced potential in two approaches of Ref.~\cite{12} and
Ref.~\cite{Ponce1and2}. However, we should remind that it has also
been shown~\cite{Freund1982} that the WEC, for $5D$ space--times,
requires $-4/3\leq\omega$, in which no other experimental
evidences have been considered.
%%%%%%%%%%%%%%%%%%%%%%%%%%%%%%%%%%%%%%%%%%%%%%%%%%%%%%%%%%%%%%%%%%%%%%%%%

%
\end{document}